\documentclass[journal = jpccck]{achemso}

\usepackage{natbib}
\usepackage{times}
\usepackage{amssymb,amsfonts,amsmath}

\newlength{\figurewidth}
\newcommand{\edit}[1]{{{ #1 }}}

\usepackage{color}

\author {Niloofar Zarifi}
\affiliation{Department of Chemistry, State University of New York at Buffalo, Buffalo, NY 14260-3000, USA}
\author {Tiange Bi}
\affiliation{Department of Chemistry, State University of New York at Buffalo, Buffalo, NY 14260-3000, USA}
\author {Hanyu Liu}
\affiliation{Innovation Center for Computational Physics Method and Software, College of Physics, Jilin University, Changchun 130012, China} 
\altaffiliation{State Key Laboratory of Superhard Materials, College of Physics, Jilin University, Changchun 130012, China}
\author {Eva Zurek}
\email{ezurek@buffalo.edu}
\affiliation{Department of Chemistry, State University of New York at Buffalo, Buffalo, NY 14260-3000, USA}

\title[An \textsf{achemso} demo]{Crystal Structures and Properties of Iron Hydrides at High Pressure}

\begin{document}

\begin{abstract}
Evolutionary algorithms and the particle swarm optimization method have been used to predict stable and metastable high hydrides of iron between 150-300~GPa that have not been discussed in previous studies. $Cmca$ FeH$_5$, $Pmma$ FeH$_6$ and $P2/c$ FeH$_6$ contain hydrogenic lattices that result from slight distortions of the previously predicted $I4/mmm$ FeH$_5$ and $Cmmm$ FeH$_6$ structures. Density functional theory calculations show that neither the $I4/mmm$ nor the $Cmca$ symmetry FeH$_5$ phases are superconducting. A $P1$ symmetry FeH$_7$ phase, which is found to be dynamically stable at 200 and 300~GPa, adds another member to the set of predicted nonmetallic transition metal hydrides under pressure. Two metastable phases of FeH$_8$ are found, and the preferred structure at 300~GPa contains a unique 1-dimensional hydrogenic lattice.  
\end{abstract}

\newpage

\section{Introduction}

The composition and structure that hydrides of iron may assume under pressure has long been of interest to geoscientists. Seismic models suggest that the Earth's core consists of iron alloyed with nickel and numerous light elements, one of which is suspected to be hydrogen \cite{Stevenson:1977}. More recently, however, it was proposed that such systems may be a route towards high density bulk atomic hydrogen \cite{Pepin2017}.  The allure of high temperature superconductivity in compressed high hydrides \cite{Shamp:2016,Zurek:2016d,Zurek:arxiv,Peng:Sc-2017} has been heightened with the discovery of superconductivity below 203~K in a sample of hydrogen sulfide that was compressed to 150~GPa \cite{Drozdov:2015a}, \edit{and most recently in studies of the lanthanum/hydrogen system \cite{la-hemley,la-eremets}}. Very high $T_c$ values have been theoretically predicted for many high hydrides including CaH$_6$ (235~K at 150~GPa) \cite{Wang:2012}, ScH$_9$ (233~K at 300~GPa) \cite{Zurek:2018b}, YH$_6$ (264~K at 120~GPa) \cite{Li:2015a}, and LaH$_{10}$ (286~K at 210~GPa) \cite{Liu:2017-La-Y}. Moreover, a number of hydrides with intriguing stoichiometries have recently been synthesized under pressure including LiH$_6$ \cite{Pepin:2015a}, NaH$_7$ \cite{Struzhkin:2016}, and LaH$_{10}$ \cite{Geballe:2018a}.

Under pressure the solubility of iron in hydrogen increases \cite{Fukai:1982} yielding phases with Fe:H ratios that approach 1:1 \cite{Badding:1991,Narygina:2011}, which undergo a number of pressure induced structural phase transitions \cite{Isaev:2007}. Evolutionary crystal structure searches coupled with density functional theory (DFT) calculations for Fe$_x$H$_y$ ($x,y=1-4$)  at 100-400~GPa found a number of unique (meta)stable structures, including FeH$_3$ in the $Pm\bar{3}m$ and $Pm\bar{3}n$ spacegroups, as well as $P2_1/m$ FeH$_4$ \cite{Bazhanova:2012a}. A further theoretical study that focused on the FeH$_4$ stoichiometry found the following sequence of phase transitions: $P2_13 \rightarrow Imma \rightarrow P2_1/m$ at 109 and 242~GPa \cite{Li:2017a}. The superconducting critical temperature, $T_c$, of the only metallic phase, $Imma$ FeH$_4$, was estimated as being 1.7~K at 110~GPa. Recently, variable-composition evolutionary searches up to 150~GPa found a number of hitherto unknown stable high hydrides of iron including $P4/mmm$ Fe$_3$H$_5$, $I4/mmm$ Fe$_3$H$_{13}$, $I4/mmm$ FeH$_5$, and FeH$_6$ with $Cmmm$ and $C2/m$ symmetries \cite{Kvashnin:2018a}. 

The theoretical study of Bazhanova and co-workers \cite{Bazhanova:2012a} inspired experiments by P\'epin et.\ al.\ that resulted in the first synthesis of the higher hydrides of iron under pressure \cite{Pepin:2014}. A ferromagnetic $I4/mmm$ FeH$_{x}$ phase with $x\sim2$ formed at 67~GPa, and at 86~GPa further hydrogen uptake yielded a nonmagnetic $Pm\bar{3}m$ FeH$_3$ phase. DFT calculations showed that both of the synthesized phases were metallic. Three years later the same group synthesized FeH$_5$ via a direct reaction between iron and H$_2$ above 130~GPa in a laser-heated diamond anvil cell \cite{Pepin2017}. Powder X-ray diffraction and Rietveld refinement determined that the lattice likely possesses $I4/mmm$ symmetry at 130~GPa. It is composed of layers of quasicubic FeH$_3$ units and atomic hydrogen whose H-H distances resembled those found in bulk atomic hydrogen \cite{Pepin2017}. DFT calculations coupled with evolutionary searches verified that $I4/mmm$ FeH$_5$ was thermodynamically stable from 85~GPa up to at least 150~GPa \cite{Kvashnin:2018a}. $I4/mmm$ FeH$_5$ was estimated to be superconducting below $\sim$50~K around 150~GPa \cite{Yansun-FeH-2017,Kvashnin:2018a}, and the $T_c$ of $Cmmm$ FeH$_6$ was estimated as being 43~K at 150~GPa \cite{Kvashnin:2018a}. However, recent calculations have questioned the conclusions of Refs.\ \cite{Yansun-FeH-2017,Kvashnin:2018a}, failing to find superconductivity within $I4/mmm$ FeH$_5$ \cite{Heil:2018a}.

Herein, we have carried out structure prediction via an evolutionary algorithm (EA) as well as the particle swarm optimization (PSO) technique to find the most stable structures of FeH$_n$ ($n = 5 - 8$) at 150, 200 and 300~GPa. A number of stable or metastable phases that have not been previously discussed are found. Importantly, in agreement with Heil et al. \cite{Heil:2018a}, but in disagreement with previous reports \cite{Yansun-FeH-2017,Kvashnin:2018a}, we do not find superconductivity in either $I4/mmm$ FeH$_5$ nor in a newly predicted $Cmca$ FeH$_5$ phase. These two dynamically stable, nearly isoenthalpic FeH$_5$ geometries are related by a slight structural distortion that decreases the $PV$ term, but increases the energetic term to the enthalpy of the $Cmca$ geometry as compared to the $I4/mmm$ arrangement. Metastable FeH$_6$, FeH$_7$ and FeH$_8$ phases that could potentially be accessed experimentally are found, and the peculiarities of their structures and electronic structures are discussed.

\section{Computational Details}
The search for stable high pressure structures with FeH$_n$ ($n = 5 - 8$) stoichiometries was carried out using the \textsc{Xtalopt} \cite{Zurek:2011a} evolutionary algorithm (EA) (releases 10 \cite{Avery:2017-xtalopt} and 11 \cite{Zurek:2017k}), and the particle swarm optimization (PSO) algorithm as implemented in the CALYPSO \cite{wang2012calypso} code. EA runs were carried out employing simulation cells with 2{-}8 formula units (FU) at 150, 200 and 300 GPa. In the EA search duplicate structures were detected via the \textsc{XtalComp} algorithm \cite{Zurek:2011i}. PSO runs were performed using  cells containing up to 4 FU at 150, 200 and 300 GPa. The lowest enthalpy structures from each search were relaxed in a pressure range from 150 to 300 GPa. 

Geometry optimizations and electronic structure calculations were performed in the framework of density functional theory (DFT) as implemented in the Vienna \textit{Ab~Initio} Simulation Package (VASP) \cite{Kresse:1993a, Kresse:1999a}  with the gradient-corrected exchange and correlation functional of Perdew{-}Burke{-}Ernzerhof (PBE) \cite{Perdew:1996a}. The projector augmented wave (PAW) method \cite{Blochl:1994a} was used to treat the core states, and a plane-wave basis set with an energy cutoff of 600 eV was employed for precise optimizations. The \edit{H 1s$^1$ and Fe 2s$^2$2p$^6$3d$^7$4s$^1$} electrons were treated explicitly in all of the calculations. The $k$-point grids were generated using the $\Gamma$-centered Monkhorst-Pack scheme, and the number of divisions along each reciprocal lattice vector was chosen such that the product of this number with the real lattice constant was 30 \AA{} in the structure searches, and 60{-}80 \AA{} otherwise. \edit{Tests carried out on $I4/mmm$ and $Cmca$ FeH$_5$ showed that the $k$-meshes and energy cutoffs used yielded relative enthalpies that were converged to within $\sim$0.1~meV/atom}. The magnetic moment was calculated for select phases. It was found to be zero, in agreement with previous studies, which found that the high hydrides of iron become nonmagnetic above 100~GPa \cite{Kvashnin:2018a}, and that the magnetic moment of $I4/mmm$ FeH$_2$ drops to zero by about 140~GPa \cite{Zhang:2018a}. 

Phonon band structures were calculated using the supercell approach \cite{parlinski1997,chaput2011phonon}, or DFT perturbation theory \cite{gonze1997dynamical}. In the former, Hellmann-Feynman forces were calculated from a supercell constructed by replicating the optimized structure wherein the atoms had been displaced, and dynamical matrices were computed using the PHONOPY code \cite{phonopy}. The Quantum Espresso (QE) \cite{giannozzi2009pwscf} program was used to obtain the dynamical matrix and electron-phonon coupling (EPC) parameters. \edit{H and Fe pseudopotentials, obtained from the QE pseudopotential library, were generated by the Vanderbilt ultrasoft method \cite{Vanderbilt:1990} with a 1s$^1$ configuration for H and a 3s$^2$3p$^6$3d$^{6.5}$4s$^1$4p$^0$ valence configuration for Fe. These are the same pseudopotentials used in Ref.\ \cite{Yansun-FeH-2017}.} The PBE generalized gradient approximation was employed. Tests were carried out to confirm that the values calculated for $I4/mmm$ FeH$_5$ at 200~GPa were in agreement with results obtained using 
PBE-PAW \edit{Troullier-Martins potentials \cite{Troullier:1991} generated by the ``atomic'' code \cite{atomic}} with valence configurations of 1s$^1$ for H and 3s$^2$3p$^6$3d$^6$4s$^2$ for Fe. The critical superconducting temperature, $T_c$, was estimated using the Allen-Dynes modified McMillan equation \cite{Allen:1975}, where the renormalized Coulomb potential, $\mu^*$, was assumed to be 0.1. Further details about the dependence of the results on the Gaussian broadening used, as well as the $k$ and $q$ meshes employed, are provided in the SI.

\section{Results and Discussion}

\subsection{Superconductivity in FeH$_5$?} 
\begin{figure}
	\begin{center}
		\includegraphics[width=0.5\columnwidth]{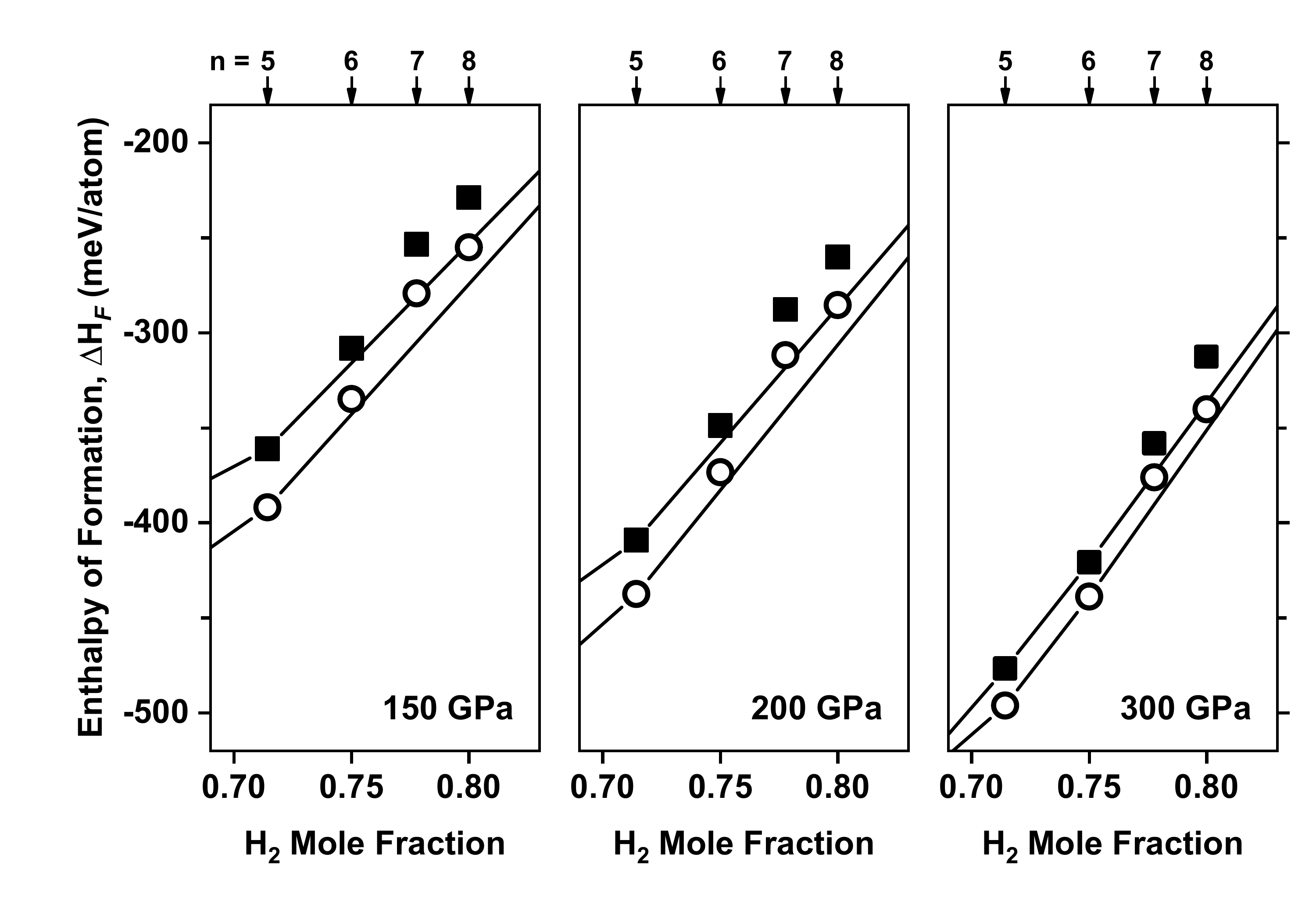}
	\end{center}
	\caption{The enthalpy of formation, $\Delta H_F$, for the reaction $\text{Fe} + \frac{n}{2} \text{H}_2 \rightarrow \text{FeH}_n$ and $n=5-8$ versus the H$_2$  mole fraction in the binary compound at different pressures. The enthalpies of $C2/c$ H$_2$ (150, 200~GPa) and $Cmca$ H$_2$ (300~GPa) \cite{Pickard-H-2007}, as well as of hcp Fe (150-300~GPa) \cite{Fe-Ref} were used to calculate $\Delta H_F$. \edit{The squares/circles denote enthalpies without/with ZPE corrections.} The lines represent the convex hull.} 
	\label{fig:chull}
\end{figure}

Fig.\ \ref{fig:chull} plots the enthalpies of formation, $\Delta H_F$, of the most stable FeH$_n$ ($n = 5-8$) compounds that were
found via crystal structure prediction (CSP) techniques at 150, 200 and 300~GPa. In a few cases, as discussed in more detail below, the enthalpies of two or more structures differed by only a few meV/atom. Phonon calculations, see Figures S3 and S4 in the SI, were carried out to confirm the dynamic stability of the predicted structures. The phases whose $\Delta H_F$ lie on the convex hull are thermodynamically stable, whereas those that do not are metastable. At all of the pressures considered only FeH$_5$ lay on the convex hull, regardless of whether the zero-point energy (ZPE) was included in the enthalpy or not. At 300~GPa FeH$_6$ lay on both hulls, but at lower pressures it was less than 9~meV/atom above the hull. At 300~GPa FeH$_7$ and FeH$_8$ lay less than 10~meV/atom above the ZPE-corrected hull, but at lower pressures the distance to the hull was larger. Because these phases are dynamically stable and they lie close to the hull, they could potentially be synthesized in experiments. \edit{For example, recently a metastable Ca$_2$H$_5$ phase that was calculated to be 20~meV/atom above the 20~GPa hull was synthesized in a laser heated diamond-anvil cell \cite{Zurek:2018c}. And, computations showed that the hydrides of phosphorus that are likely contributors to the superconductivity measured in compressed phosphine between  $\sim$80-225~GPa \cite{Drozdov:2015-P} are dynamically stable, but unstable with respect to decomposition into the elemental phases by at least 30~meV/atom \cite{Zurek:2015j,Zurek:2017c}.}  

\begin{figure}
	\begin{center}
		\includegraphics[width=0.5\columnwidth]{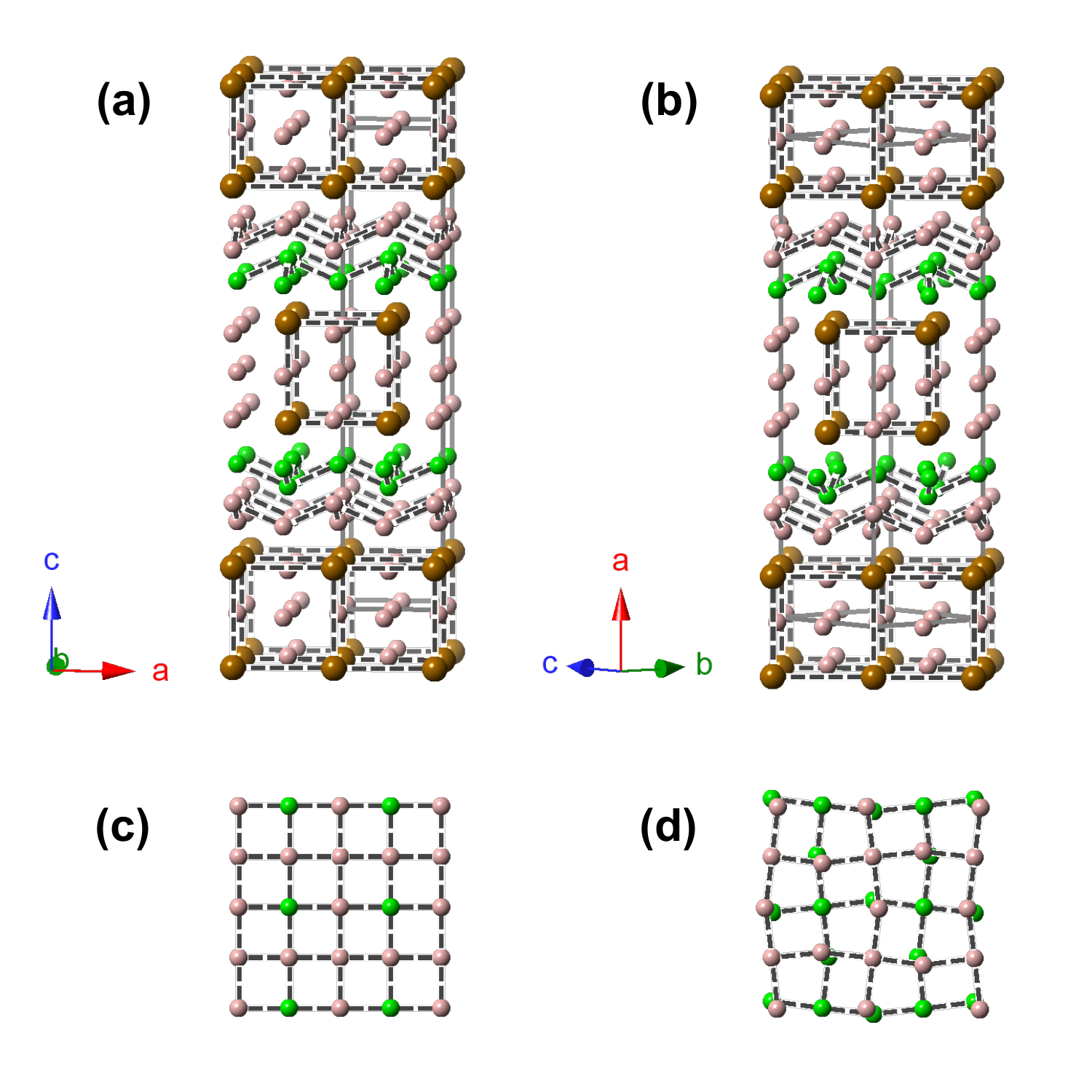}
	\end{center}
	\caption{Crystal structures of (a) $I4/mmm$ FeH$_5$ and (b) $Cmca$ FeH$_5$ at 300~GPa. Fe atoms are shown in brown and hydrogen atoms in pink and green. The FeH$_3$ unit that is found in a number of high-pressure iron hydrides is outlined by the dashed black lines. Views of the puckered hydrogen layers in (c) $I4/mmm$ FeH$_5$ projected onto the $ab$ plane, and (d) $Cmca$ FeH$_5$ projected onto the $bc$ plane are also provided. Connections are drawn between H-H and Fe-Fe nearest neighbor atoms to more clearly illustrate the structure, and they are not indicative of bonds.} 
	\label{fig:FeH5}
\end{figure}

At 150 and 200~GPa the most stable FeH$_5$ structure identified in our searches possessed $I4/mmm$ symmetry, in agreement with the phase that was proposed experimentally \cite{Pepin2017}. Recently, this same phase was found using various CSP techniques, and it was computed to be thermodynamically stable from 85-300~GPa \cite{Kvashnin:2018a,Zhang:2018a}. At 300~GPa our searches also revealed a $Cmca$ structure that was nearly isoenthalpic with $I4/mmm$ FeH$_5$. The enthalpies of these two phases differed by less than 1~meV/atom, even when the ZPE was taken into account. Both crystals, illustrated in Fig.\ \ref{fig:FeH5}, are comprised of face-sharing simple-cubic like layers of FeH$_3$ units, which resemble the high pressure $Pm\bar{3}m$ FeH$_3$ phase synthesized in Ref.\ \cite{Pepin:2014}, arranged in an ABAB...\ stacking. A number of previously predicted high pressure iron hydride phases contained this FeH$_3$ motif, including $I4/mmm$ FeH$_2$, $I4/mmm$ Fe$_3$H$_{13}$ and $Cmmm$ FeH$_6$ \cite{Kvashnin:2018a}. In FeH$_5$ these units are separated by layers of hydrogen atoms that form puckered hexagonal honeycomb sheets. At 300~GPa the nearest neighbor Fe-Fe, Fe-H and H-H distances are nearly the same in the two phases: $2.238$/$2.239$~\AA{}, $1.426$/$1.424$~\AA{}, and $1.232$/$1.236$~\AA{}, respectively, in $I4/mmm$~/~$Cmca$. Plots of the electron localization function (ELF) do not provide any evidence of covalent bond formation. For example, the ELF at the midpoint between two H atoms is $\sim$0.55, in-line with the long H-H distances.

A unit cell of the $Cmca$ structure contains twice as many atoms as does $I4/mmm$. However, the main difference between these phases originates from their two stacked hydrogenic layers, colored green and pink in Fig.\ \ref{fig:FeH5}. When projected onto the $ab$ plane in the $I4/mmm$ lattice the H-H-H angles measure 90$^\circ$ and 180$^\circ$, whereas a similar projection in the $Cmma$ structure reveals a distortion has taken place resulting in smaller H-H-H angles and a doubling of the unit cell. As shown in the SI, this structural distortion yields a slightly smaller volume for the $Cmca$ phase above $\sim$225~GPa, giving rise to a smaller $PV$ contribution to the enthalpy. The electronic contribution, on the other hand, favors the $I4/mmm$ structure. These opposing effects cancel each other out, so the two phases are nearly isoenthalpic. Similar behavior has been observed for the $I4/mmm$ and $C2/m$ PH$_2$ structures \cite{Zurek:2015j} that were proposed to contribute to the superconductivity observed when phosphine was compressed to 207~GPa \cite{Drozdov:2015-P}. 

\begin{figure}
 	\begin{center}
 		\includegraphics[width=0.5\columnwidth]{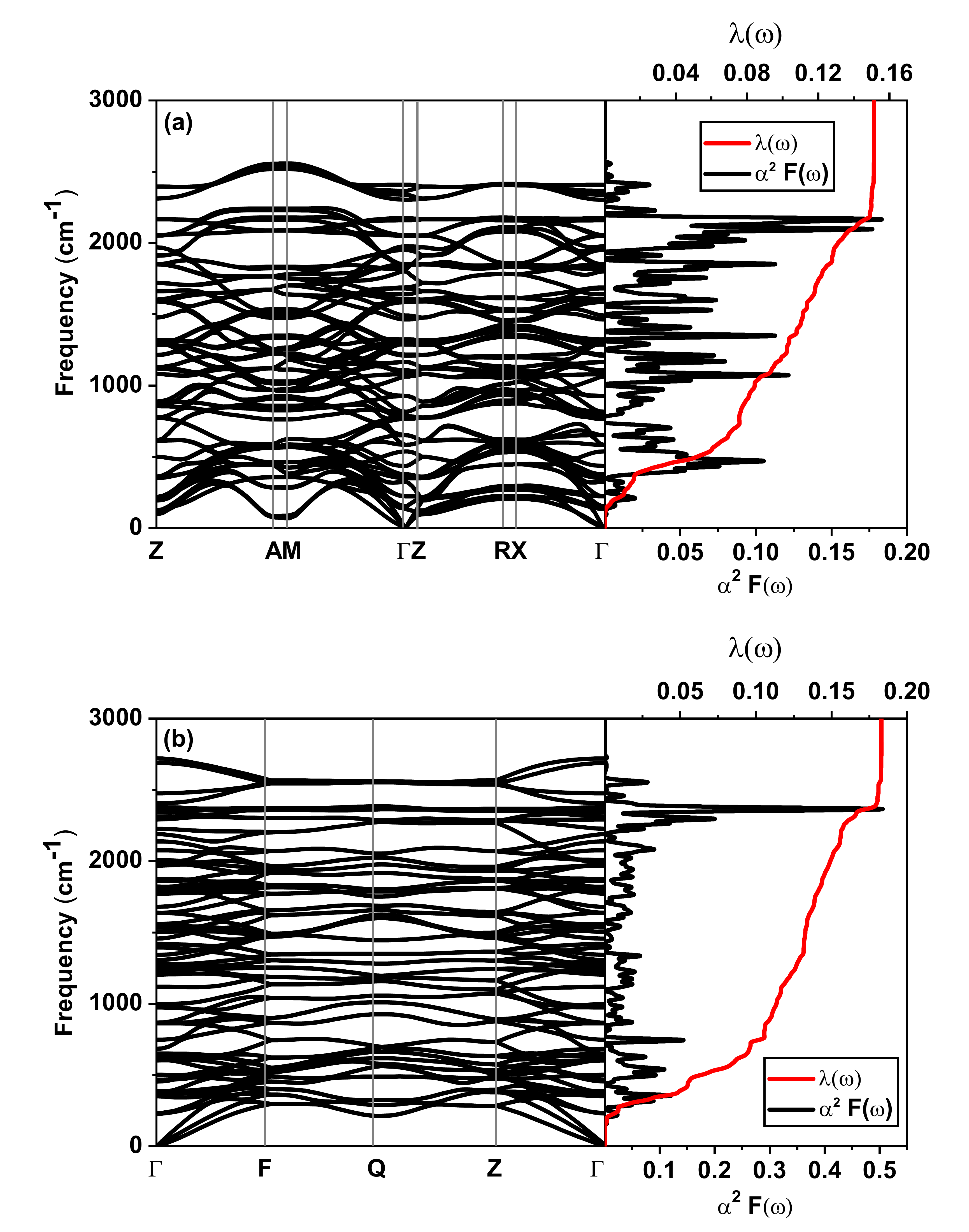}
 	\end{center}
 	\caption{Phonon band structures, the Eliashberg spectral function, $\alpha^2F(\omega)$, and the electron-phonon integral, $\lambda(\omega)$, for (a) $I4/mmm$ FeH$_5$ at 200~GPa, and (b) $Cmca$ FeH$_5$ at 300~GPa.
 	} 
 	\label{fig:Tc}
 \end{figure}

It was pointed out \cite{Yansun-FeH-2017} that the hydrogenic layers in $I4/mmm$ FeH$_5$ resemble the $Cmca$ phase of hydrogen that was predicted to be stable between 385-490~GPa \cite{Pickard-H-2007}. SC-DFT calculations yielded a $T_c$ of 242~K at 450~GPa for this H$_2$ phase \cite{Cudazzo:2008a}. Its nearest neighbor H-H distances, 0.80~\AA{}, are significantly shorter than those found within FeH$_5$ in the pressure range considered here. Despite this, in Ref.\ \cite{Yansun-FeH-2017} it was postulated that the structural similarities between $Cmca$ H$_2$ and the hydrogenic layers in $I4/mmm$ FeH$_5$ would render the latter phase superconducting. The electron phonon coupling, $\lambda$, and logarithmic average phonon frequency $\omega_\text{log}$  was calculated to be 1.13 and 614~K at 130~GPa \cite{Yansun-FeH-2017}, and 0.97 and 642.3~K at 150~GPa \cite{Kvashnin:2018a}. Using a value of 0.1 for the Coulomb pseudopotential, $\mu^*$, $T_c$ was estimated as being 51/46~K at 130/150~GPa via the Allen-Dynes equation, and 43~K at 150~GPa via the McMillan equation \cite{Yansun-FeH-2017,Kvashnin:2018a}. On the other hand, recently Heil, Bachelet and Boeri used Migdal-Eliashberg theory to show that the $T_c$ of FeH$_5$ at these conditions is actually $\le$1~K \cite{Heil:2018a}. Our calculations (details are provided in the SI) are in good agreement with those of Heil et.\ al. Using the \textsc{Quantum Espresso} \cite{giannozzi2009pwscf} package we obtain $\lambda=$~0.15/0.18, $\omega_\text{log}=$~1130/1085~K for $I4/mmm$ / $Cmca$ FeH$_5$ at 200/300~GPa, both yielding a $T_c<1$~K for $\mu^*=0.1$. The computed phonon band structure and Eliashberg spectral function for these phases is provided in Fig.\ \ref{fig:Tc}. For comparison, at 150~GPa Heil and co-workers obtained $\lambda=$~0.14 and $\omega_\text{log}=$~1050~K yielding a $T_c$ of ~0~K for $\mu^*=0.16$ via a more elaborate calculation carried out using the \textsc{EPW} code \cite{EPW}. 

The calculation of $\lambda$ within \textsc{Quantum Espresso} requires a double-delta integration over the Fermi surface \cite{Wierzbowska} (see the SI for further details). This integration can be performed by using very dense $k$-point (electronic) and $q$-point (phonon wave-vector) grids where the $\delta$ functions are approximated using Gaussians. In studies of solid atomic hydrogen at high pressures the broadenings that yielded converged $\lambda$ were typically between 0.02 to 0.025~Ry., but some phases required broadenings of 0.035~Ry.\ \cite{Mcmahon-2011}. In our work $\lambda$ was converged for broadenings of 0.04-0.05~Ry. In the SI we illustrate that $T_c$ values in-line with those calculated in Refs.\ \cite{Yansun-FeH-2017,Kvashnin:2018a} can be obtained within \textsc{Quantum Espresso} using very large Gaussian broadenings of $\sim$0.275~Ry. 

First principles calculations did not find superconductivity in $Pm\bar{3}m$ FeH$_3$ at 150~GPa \cite{Heil:2018a}, nor in $Fm\bar{3}m$ CoH$_2$, $I4/mmm$ CoH$_2$ and $Pm\bar{3}m$ CoH$_3$ up to 200~GPa \cite{CoH-Cui-2018} (the latter two structures are isotypic with the previously synthesized iron analogues \cite{Pepin:2014}). The maximum $T_c$s for the predicted $Fm\bar{3}m$ RuH, $Pm\bar{3}m$ RuH$_3$ and $Pm\bar{3}n$ RuH$_3$ phases were calculated to be 0.4~K (100~GPa), 3.6~K (100~GPa) and 1.3~K (200~GPa) at the pressures given in the parentheses \cite{Liu:2015c}. And a $T_c$ of 2.1~K at 100~GPa was calculated for an $Fm\bar{3}m$ symmetry OsH phase \cite{Liu:2015b}. These estimates do not bode well for superconductivity within the Group 8 polyhydrides under pressure.

\subsection{Structural Distortions in FeH$_6$} 
\begin{figure*}
 	\begin{center}
 		\includegraphics[width=0.9\columnwidth]{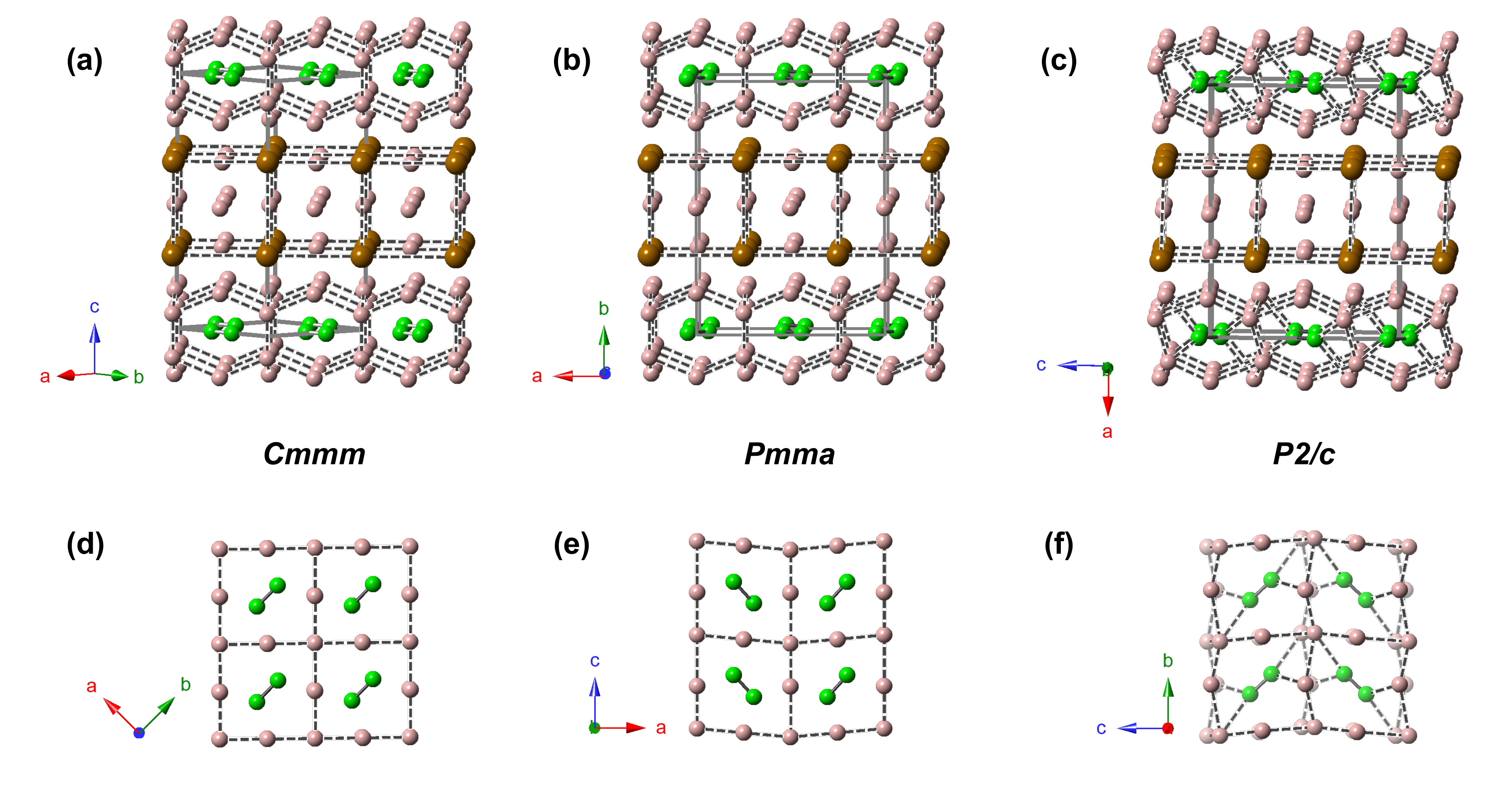}
 	\end{center}
 	\caption{Crystal structures of (a) $Cmmm$, (b) $Pmma$ and (c) $P2/c$ FeH$_6$ at 200~GPa. The FeH$_3$ unit that is found in a number of high-pressure iron hydrides is outlined by the dashed black lines. (d,e,f) A top-down view of the hydrogen layers is provided below each structure to more clearly illustrate the differences between them. Monoatomic hydrogen atoms are colored pink and hydrogen atoms belonging to H$_2$ are colored green. 
 	} 
 	\label{fig:FeH6-200}
\end{figure*}

For the FeH$_6$ stoichiometry evolutionary searches predicted that a $C2/m$ symmetry phase would be thermodynamically stable between 35-82~GPa \cite{Kvashnin:2018a}. Above this pressure it was found to transform to the $Cmmm$ phase illustrated in Fig.\ \ref{fig:FeH6-200}(a), which lay on the convex hull up to 115~GPa and was only 1.5~meV/atom above the hull up to 150~GPa. PSO searches, on the other hand, found that $C2/m$ FeH$_6$ transformed into a $Cmcm$ symmetry structure at 100~GPa, followed by the $Cmmm$ phase in Fig.\ \ref{fig:FeH6-200}(a), which was computed to be stable between 106.8 - 115~GPa \cite{Zhang:2018a}. This same study concluded that the FeH$_6$ stoichiometry is not thermodynamically stable between 115-213.7~GPa, but above this pressure a nonmetallic $C2/c$ FeH$_6$ phase became stable up to at least 300~GPa. In addition to $Cmmm$ FeH$_6$, our EA searches found the $Pmma$ and $P2/c$ phases illustrated in Fig.\ \ref{fig:FeH6-200} (b, c). \edit{Not taking into account the ZPE, the enthalpies of these three phases were within 1~meV/atom of each other up to 180~GPa.} When the ZPE was included, the enthalpy of the $Pmma$ phase was $\sim$2~meV/atom lower than $P2/c$ at 150~GPa. Above 200~GPa the $C2/c$ phase found previously in Ref.\ \cite{Zhang:2018a} became preferred up to the highest pressures considered herein, and it lay on the 300~GPa hull. 

The $Cmmm$, $Pmma$ and $P2/c$ symmetry FeH$_6$ structures are all related, since they are comprised of layers of the same face-sharing FeH$_3$ units found in FeH$_5$ separated by two puckered layers of hydrogen atoms that are in turn separated by a layer of H$_2$ molecules. The main difference between these three phases arises from small distortions of their hydrogenic sublattices. As shown in Fig.\ \ref{fig:FeH6-200}\edit{(d)}, when the hydrogen atoms in the $Cmmm$ phase are projected onto the $ab$ plane the H-H-H angles measure 90.7$^\circ$ and 89.3$^\circ$. In the $Pmma$ structure these angles measure 95.3$^\circ$ and 84.7$^\circ$ instead. Moreover, whereas the H$_2$ molecules are parallel to each other in the $Cmmm$ phase, in $Pmma$ every second H$_2$ molecule has been rotated by 90$^\circ$ along the $a$-axis, \edit{as evident from Fig.\ \ref{fig:FeH6-200}(e)}. As shown in Figure S5, these slight distortions result in a smaller volume for the $Pmma$ phase in the whole pressure range studied. As the pressure increases, so does the difference in the volume and the $PV$ contribution to the enthalpy favors the more compact $Pmma$ phase. The electronic contribution to the enthalpy for the two phases is nearly the same at 150~GPa, but because of the $PV$ term $Pmma$ FeH$_6$ is favored by \edit{0.6~meV/atom} at this pressure.

The $P2/c$ phase represents a further distortion of the hydrogenic sublattice. The main difference between the $Pmma$ and $P2/c$ geometries is that the two layers of hydrogen atoms are no longer stacked on top of each other in the latter. The shear in the lattice, which can be seen most clearly when comparing Fig.\ \ref{fig:FeH6-200}(e) with Fig.\ \ref{fig:FeH6-200}(f), results in a smaller volume in the $P2/c$ phase above 160~GPa. The smaller $PV$ contribution to the enthalpy is the driving force for the stabilization of this phase \edit{by 0.2~meV/atom at 200~GPa}. At 200~GPa the H-H bond lengths in the three phases are nearly equidistant, measuring 0.729~\AA{}, 0.730~\AA{} and 0.733~\AA{} in $Cmmm$, $Pmma$ and $C2/c$, respectively. Phonon calculations revealed that $Cmmm$ FeH$_6$ is dynamically stable only at 100~GPa, whereas $Pmma$ FeH$_6$ is dynamically stable at 150~GPa, and $P2/c$ FeH$_6$ is dynamically stable at 150~GPa and 200~GPa.

Above 200~GPa $P2/c$ FeH$_6$ becomes unstable with respect to the $C2/c$ phase that was predicted via the PSO technique \cite{Zhang:2018a}. This structure does not possess any molecular hydrogen units, with the shortest H-H distance measuring 1.156~\AA{} at 300~GPa. The crystal structure of this phase, along with its remarkable electronic structure that renders it nonmetallic, has been described fully in Ref.\ \cite{Zhang:2018a}.

\subsection{Metastable FeH$_7$ and FeH$_8$}     
\begin{figure}
\begin{center}
\includegraphics[width=0.45\columnwidth]{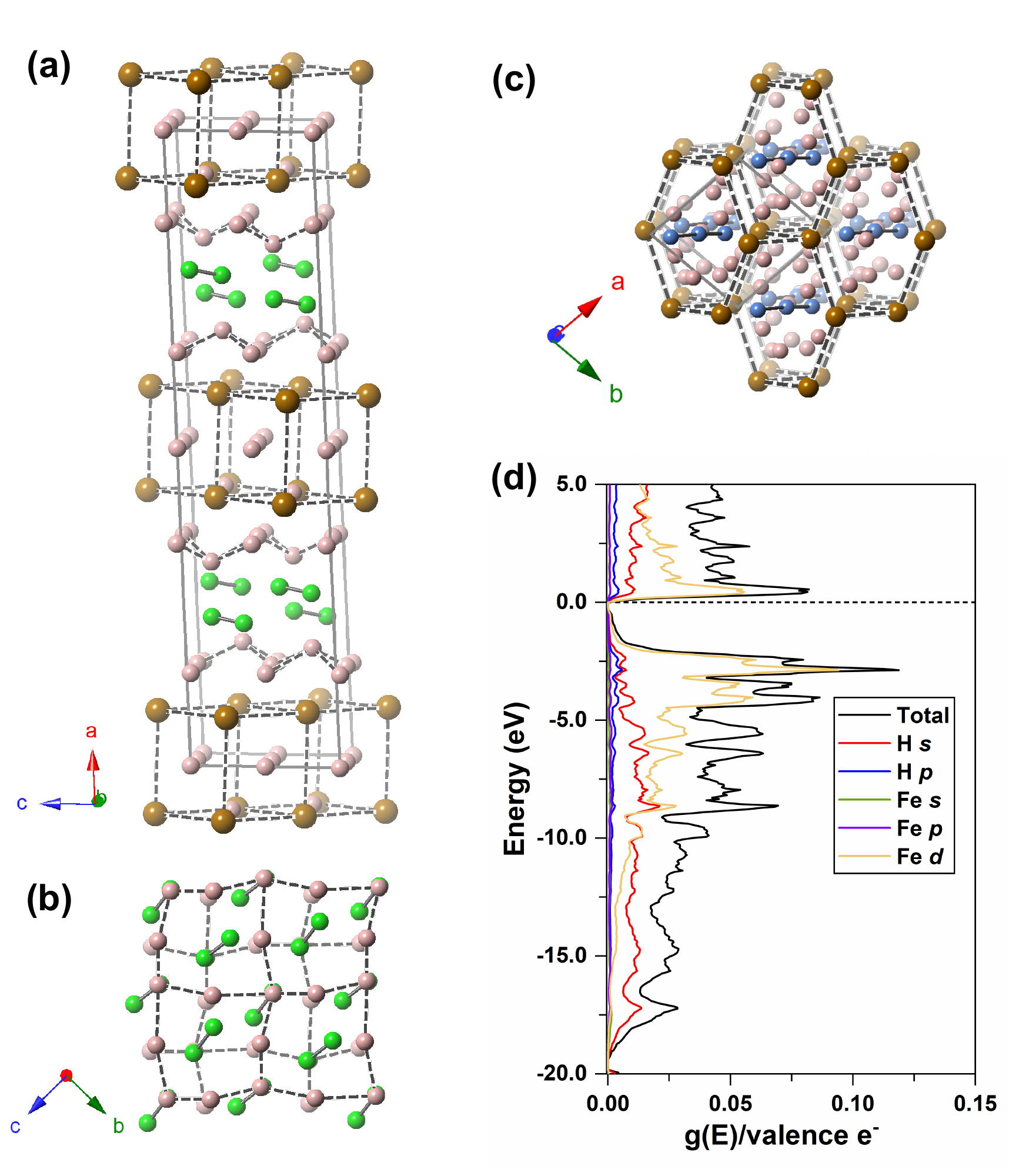}
\end{center}
\caption{(a) Crystal structure of the $C$2/$c$ FeH$_7$ phase at 150~GPa, and (b) a view of its hydrogenic layers projected onto the $bc$ plane. (c) Crystal structure, and (d) total and projected DOS of $P1$ FeH$_7$ at 300~GPa. Monoatomic hydrogens are colored pink, hydrogen atoms belonging to H$_2$ units are colored green, and hydrogen atoms belonging to H$_3^-$ units are colored purple.}  
\label{fig:FeH7}
\end{figure}

At 150~GPa the dynamically stable $C2/c$ FeH$_7$ phase illustrated in Fig.\ \ref{fig:FeH7}(a) had the lowest enthalpy and it remained the most stable geometry up to 190~GPa. Like FeH$_5$ and FeH$_6$, this phase is comprised of layers of face-sharing FeH$_3$ units that assume an ABAB... stacking. Similar to FeH$_6$, the FeH$_3$ slabs are separated by puckered layers comprised of hydrogen atoms and layers of H$_2$ molecules whose H-H distance measures 0.759~\AA{}. A comparison of the hydrogenic layers of $P2/c$ FeH$_6$ and $C2/c$ FeH$_7$, shown in Fig.\ \ref{fig:FeH6-200}(f) and Fig.\ \ref{fig:FeH7}(b), suggests the two may be related by a structural distortion.

Above 190~GPa, the $P1$ symmetry FeH$_7$ phase illustrated in Fig.\ \ref{fig:FeH7}(c) was preferred. Phonon calculations showed this phase was dynamically stable at 200 and 300~GPa. Each iron atom in the three-dimensional Fe framework was coordinated to four others with Fe-Fe distances of 2.28-2.34~\AA{} at 300~GPa, which is somewhat shorter than the 2.48~\AA{} found for the $Im\bar{3}m$ Fe phase at ambient conditions \cite{Kohlhaas:1967}. The iron sublattice of $P1$ FeH$_7$ resembled that of the previously predicted $C2/c$ FeH$_6$ phase \cite{Zhang:2018a}, with the main difference between them being the shape of the channels that run along the $c$-axis. In both systems hydrogen lies within these channels, however whereas $C2/c$ FeH$_6$ is comprised only of hydrogen atoms, $P1$ FeH$_7$ contains both hydrogen atoms and H$_3^-$-like units with H-H distances measuring 0.894 and 0.986~\AA{}, and an H-H-H angle of 176.3$^{\circ}$. Such linear H$_3^-$ motifs have been previously predicted in numerous polyhydride phases under pressure \cite{Zurek:2011h,Zhong:2016a}. The shape adopted by the Fe framework, and the somewhat longer Fe-Fe distances in FeH$_7$ as compared to FeH$_6$, can be attributed to the structural nuances of the hydrogenic lattice. It is interesting to note that like $C2/c$ FeH$_6$, $P1$ FeH$_7$ is nonmetallic, as is evident in the plot of the density of states (DOS) in Fig.\ \ref{fig:FeH7}(d). Within PBE the band gap of the former is computed to be 0.38~eV at 300~GPa (cf.\ 0.93~eV at 220~GPa with the HSE06 functional \cite{Zhang:2018a}), whereas we calculate the PBE band gap of the latter to be 0.32~eV at 200~GPa and 0.11~eV at 300~GPa. Thus, in addition to $P2_1/m$ FeH$_4$ \cite{Li:2017a}, and $C2/c$ FeH$_6$ \cite{Zhang:2018a}, $P1$ FeH$_7$ is another nonmetallic transition metal hydride at high pressure.

\begin{figure}
\begin{center}
\includegraphics[width=0.7\columnwidth]{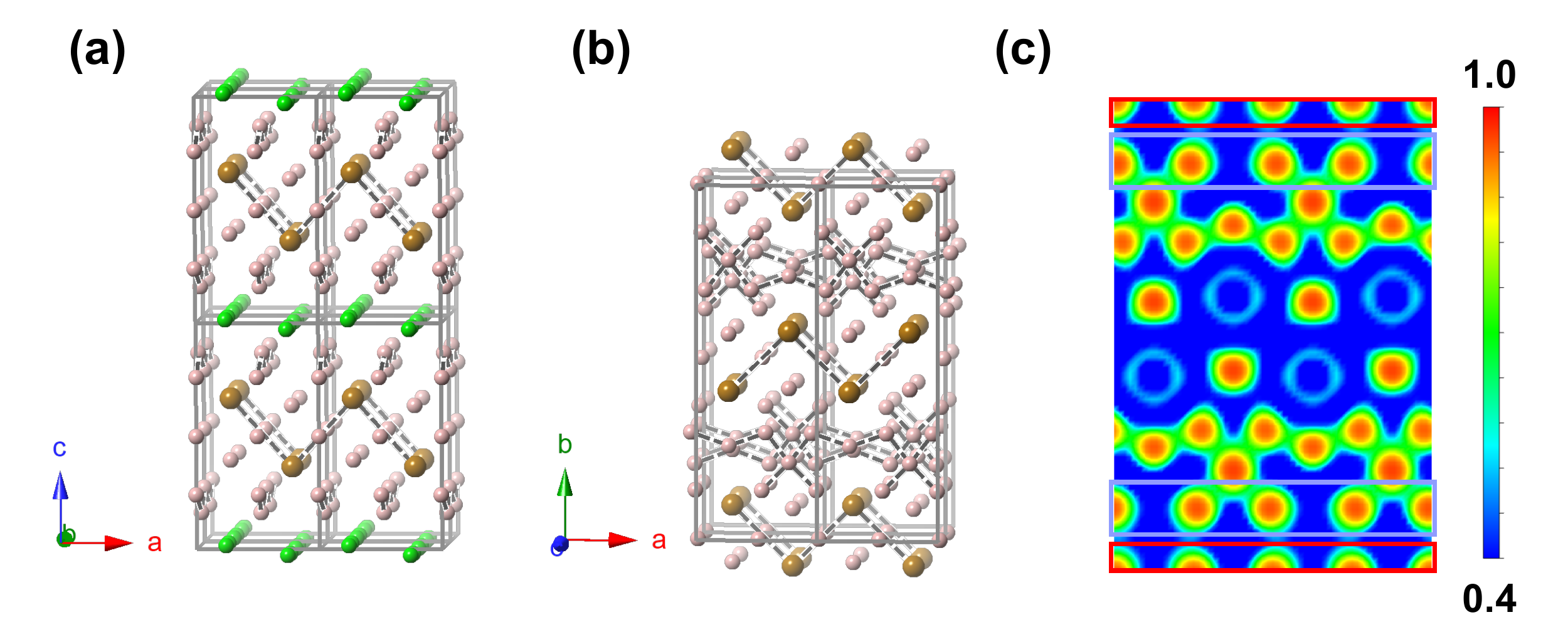}
\end{center}
\caption{Crystal structures of (a) $Pmma$ FeH$_8$ at 150~GPa, and (b) $Ima2$ FeH$_8$ at 300~GPa. (c) A plot of the ELF of $Ima2$ FeH$_8$ viewed in the $ab$ plane with $z=$0.742. } \label{fig:FeH8}
\end{figure}

The most stable FeH$_8$ phase found possessed $Pmma$ symmetry up to 290~GPa and assumed $Ima2$ symmetry at higher pressures. As illustrated in Fig.\ \ref{fig:FeH8}(a), it consisted of zig-zag motifs that could be constructed by removing atoms from the FeH$_3$ building blocks common to many of the other high pressure iron hydrides, separated by layers of atomic hydrogen and H$_2$ molecules whose H-H distance measured 0.76~\AA{} at 150~GPa. Phonon calculations revealed this phase was dynamically stable at 150 and 200~GPa. The $Ima2$ phase, shown in Fig.\ \ref{fig:FeH8}(b) is also comprised of these same zig-zag motifs, but they are arranged in an ABAB... stacking. At 300~GPa 1-dimensional hydrogenic motifs that have not been observed in any other high pressure polyhydride emerged in this phase. In these chains, which resembled a row of ``X'' motifs joined by one hydrogen atom, the H-H distances ranged from 0.99-1.06~\AA{}, resulting in a bonding interaction, as shown via the plot of the ELF in Fig.\ \ref{fig:FeH8}(c). Note that the nearest neighbor distance between the hydrogen atoms outlined in the red and purple boxes in Fig.\ \ref{fig:FeH8}(c) was 1.290-1.322~\AA{}, so there was no bonding interaction between them.

\section{Conclusions}

Crystal structure prediction techniques coupled with density functional theory calculations were employed to find the most stable FeH$_n$ ($n=5-8$) structures at 150, 200 and 300~GPa. The geometries, electronic structures and propensity for superconductivity of stable and metastable phases has been investigated. Our work adds the following contributions to the already published theoretical studies of hydrides of iron under pressure \cite{Bazhanova:2012a,Yansun-FeH-2017,Kvashnin:2018a,Li:2017a,Zhang:2018a}:

\begin{itemize}
 \item in disagreement with the results of Majumdar et al.\ \cite{Yansun-FeH-2017} and Kvashnin et al.\ \cite{Kvashnin:2018a}, but in agreement with Heil and co-workers \cite{Heil:2018a} we do not find superconductivity in the recently synthesized $I4/mmm$ FeH$_5$ structure \cite{Pepin2017}, nor in the structurally similar and isoenthalpic $Cmca$ FeH$_5$ phase.
 \item via crystal structure prediction techniques we have found three new phases, $Cmca$ FeH$_5$, $Pmma$ FeH$_6$, and $P2/c$ FeH$_6$, whose hydrogenic lattices can be derived by distortions of the previously reported $I4/mmm$ FeH$_5$ and $Cmmm$ FeH$_6$ phases. The structural distortions decrease the volume and the $PV$ contribution to the enthalpy, however they increase the electronic energy. The interplay between these two factors can render two or more distinct, dynamically stable phases nearly isoenthalpic within a given pressure range. Increasing pressure typically tends to stabilize the phase with the smaller volume.
 \item we add one more nonmetallic high pressure transition metal hydride, $P1$ FeH$_7$, to the list of previously reported insulating structures: $P2_13$ and $P2_1/m$ FeH$_4$ \cite{Li:2017a}, and $C2/c$ FeH$_6$ \cite{Zhang:2018a}. $P1$ FeH$_7$, whose structure resembles that of $C2/c$ FeH$_6$,  is dynamically stable at 200 and 300~GPa. At these pressures it is sufficiently close to the convex hull so that it could potentially be synthesized in experiments. 
 \item two metastable FeH$_8$ phases are predicted, with the structure found at 150 and 200~GPa being composed of motifs that could be derived from the FeH$_3$ building blocks common to many of the high pressure iron hydrides. The 300~GPa structure contains a 1-dimensional hydrogenic sublattice that has not been previously observed.
\end{itemize}

\section*{Supporting Information}
The Supporting Information is available free of charge on the ACS Publications website at DOI: xxx. The Supporting Information includes: structural parameters, equations of state of select phases, projected densities of states, phonon band structures, dependence of $\lambda$, $\omega_\text{log}$, and $T_c$ on the parameters employed in the QE calculations.

\section*{Acknowledgements}
We acknowledge the NSF (DMR-1505817) for financial, and the Center for Computational Research (CCR) at SUNY Buffalo for computational support.


\begin{mcitethebibliography}{59}
\providecommand*\natexlab[1]{#1}
\providecommand*\mciteSetBstSublistMode[1]{}
\providecommand*\mciteSetBstMaxWidthForm[2]{}
\providecommand*\mciteBstWouldAddEndPuncttrue
  {\def\EndOfBibitem{\unskip.}}
\providecommand*\mciteBstWouldAddEndPunctfalse
  {\let\EndOfBibitem\relax}
\providecommand*\mciteSetBstMidEndSepPunct[3]{}
\providecommand*\mciteSetBstSublistLabelBeginEnd[3]{}
\providecommand*\EndOfBibitem{}
\mciteSetBstSublistMode{f}
\mciteSetBstMaxWidthForm{subitem}{(\alph{mcitesubitemcount})}
\mciteSetBstSublistLabelBeginEnd
  {\mcitemaxwidthsubitemform\space}
  {\relax}
  {\relax}

\bibitem[Stevenson(1977)]{Stevenson:1977}
Stevenson,~D.~J. Hydrogen in the Earth's Core. \emph{Nature} \textbf{1977},
  \emph{268}, 130--131\relax
\mciteBstWouldAddEndPuncttrue
\mciteSetBstMidEndSepPunct{\mcitedefaultmidpunct}
{\mcitedefaultendpunct}{\mcitedefaultseppunct}\relax
\EndOfBibitem
\bibitem[P{\'e}pin et~al.(2017)P{\'e}pin, Geneste, Dewaele, Mezouar, and
  Loubeyre]{Pepin2017}
P{\'e}pin,~C.~M.; Geneste,~G.; Dewaele,~A.; Mezouar,~M.; Loubeyre,~P. Synthesis
  of FeH$_5$: A Layered Structure with Atomic Hydrogen Slabs. \emph{Science}
  \textbf{2017}, \emph{357}, 382--385\relax
\mciteBstWouldAddEndPuncttrue
\mciteSetBstMidEndSepPunct{\mcitedefaultmidpunct}
{\mcitedefaultendpunct}{\mcitedefaultseppunct}\relax
\EndOfBibitem
\bibitem[Shamp and Zurek(2017)Shamp, and Zurek]{Shamp:2016}
Shamp,~A.; Zurek,~E. Superconductivity in Hydrides Doped with Main Group
  Elements Under Pressure. \emph{Nov. Supercond. Mater.} \textbf{2017},
  \emph{3}, 14--22\relax
\mciteBstWouldAddEndPuncttrue
\mciteSetBstMidEndSepPunct{\mcitedefaultmidpunct}
{\mcitedefaultendpunct}{\mcitedefaultseppunct}\relax
\EndOfBibitem
\bibitem[Zurek(2017)]{Zurek:2016d}
Zurek,~E. Hydrides of the Alkali Metals and Alkaline Earth Metals Under
  Pressure. \emph{Comments Inorg. Chem.} \textbf{2017}, \emph{37}, 78--98\relax
\mciteBstWouldAddEndPuncttrue
\mciteSetBstMidEndSepPunct{\mcitedefaultmidpunct}
{\mcitedefaultendpunct}{\mcitedefaultseppunct}\relax
\EndOfBibitem
\bibitem[Bi et~al.(2018)Bi, Zarifi, Terpstra, and Zurek]{Zurek:arxiv}
Bi,~T.; Zarifi,~N.; Terpstra,~T.; Zurek,~E. The Search for Superconductivity in
  High Pressure Hydrides. \emph{arXiv preprint} \textbf{2018},
  arXiv:1806.00163\relax
\mciteBstWouldAddEndPuncttrue
\mciteSetBstMidEndSepPunct{\mcitedefaultmidpunct}
{\mcitedefaultendpunct}{\mcitedefaultseppunct}\relax
\EndOfBibitem
\bibitem[Peng et~al.(2017)Peng, Sun, Pickard, Needs, Wu, and Ma]{Peng:Sc-2017}
Peng,~F.; Sun,~Y.; Pickard,~C.~J.; Needs,~R.~J.; Wu,~Q.; Ma,~Y. Hydrogen
  Clathrate Structures in Rare Earth Hydrides at High Pressures: Possible Route
  to Room-Temperature Superconductivity. \emph{Phys. Rev. Lett.} \textbf{2017},
  \emph{119}, 107001 (1--6)\relax
\mciteBstWouldAddEndPuncttrue
\mciteSetBstMidEndSepPunct{\mcitedefaultmidpunct}
{\mcitedefaultendpunct}{\mcitedefaultseppunct}\relax
\EndOfBibitem
\bibitem[Drozdov et~al.(2015)Drozdov, Eremets, Troyan, Ksenofontov, and
  Shylin]{Drozdov:2015a}
Drozdov,~A.~P.; Eremets,~M.~I.; Troyan,~I.~A.; Ksenofontov,~V.; Shylin,~S.~I.
  Conventional Superconductivity at 203~Kelvin at High Pressures in the Sulfur
  Hydride System. \emph{Nature} \textbf{2015}, \emph{525}, 73--76\relax
\mciteBstWouldAddEndPuncttrue
\mciteSetBstMidEndSepPunct{\mcitedefaultmidpunct}
{\mcitedefaultendpunct}{\mcitedefaultseppunct}\relax
\EndOfBibitem
\bibitem[Somayazulu et~al.(2018)Somayazulu, Ahart, Mishra, Geballe, Baldini, Meng, Struzhkin, and Hemley]{la-hemley}
Somayazulu,~M.; Ahart,~M.; Mishra,~A.~K.; Geballe,~Z.~M.; Baldini,~M.; Meng,~Y.; Struzhkin,~V.~V.; Hemley,~R.~J.
  Evidence for Superconductivity above 260~K in Lanthanum Superhydride at Megabar Pressures. \emph{arXiv preprint} \textbf{2018},
  arXiv:1808.07695\relax
\mciteBstWouldAddEndPuncttrue
\mciteSetBstMidEndSepPunct{\mcitedefaultmidpunct}
{\mcitedefaultendpunct}{\mcitedefaultseppunct}\relax
\EndOfBibitem
\bibitem[Drozdov et~al.(2018)Drozdov, Minkov, Besedin, Kong, Kuzovnikov, Knyazev, and Eremets]{la-eremets}
Drozdov,~A.~P.; Minkov,~V.~S.; Besedin,~S.~P.; Kong,~P.~P.; Kuzovnikov,~M.~A.; Knyazev,~D.~A.; Eremets, M. I.
  Superconductivity at 215~K in Lanthanum Hydride at High Pressures. \emph{arXiv preprint} \textbf{2018},
  arXiv:1808.07039\relax
\mciteBstWouldAddEndPuncttrue
\mciteSetBstMidEndSepPunct{\mcitedefaultmidpunct}
{\mcitedefaultendpunct}{\mcitedefaultseppunct}\relax
\EndOfBibitem
\bibitem[Wang et~al.(2012)Wang, Tse, Tanaka, Iitaka, and Ma]{Wang:2012}
Wang,~H.; Tse,~J.~S.; Tanaka,~K.; Iitaka,~T.; Ma,~Y. Superconductive
  Sodalite-like Clathrate Calcium Hydride at High Pressures. \emph{Proc. Natl.
  Acad. Sci. U.S.A.} \textbf{2012}, \emph{109}, 6463--6466\relax
\mciteBstWouldAddEndPuncttrue
\mciteSetBstMidEndSepPunct{\mcitedefaultmidpunct}
{\mcitedefaultendpunct}{\mcitedefaultseppunct}\relax
\EndOfBibitem
\bibitem[Ye et~al.(2018)Ye, Zarifi, Zurek, Hoffmann, and Ashcroft]{Zurek:2018b}
Ye,~X.; Zarifi,~N.; Zurek,~E.; Hoffmann,~R.; Ashcroft,~N.~W. High Hydrides of
  Scandium under Pressure: Potential Superconductors. \emph{J. Phys. Chem. C}
  \textbf{2018}, \emph{122}, 6298--6309\relax
\mciteBstWouldAddEndPuncttrue
\mciteSetBstMidEndSepPunct{\mcitedefaultmidpunct}
{\mcitedefaultendpunct}{\mcitedefaultseppunct}\relax
\EndOfBibitem
\bibitem[Li et~al.(2015)Li, Hao, Liu, Tse, Wang, and Ma]{Li:2015a}
Li,~Y.; Hao,~J.; Liu,~H.; Tse,~J.~S.; Wang,~Y.; Ma,~Y. Pressure-Stabilized
  Superconductive Yttrium Hydrides. \emph{Sci. Rep.} \textbf{2015}, \emph{5},
  9948 (1--8)\relax
\mciteBstWouldAddEndPuncttrue
\mciteSetBstMidEndSepPunct{\mcitedefaultmidpunct}
{\mcitedefaultendpunct}{\mcitedefaultseppunct}\relax
\EndOfBibitem
\bibitem[Liu et~al.(2017)Liu, Naumov, Hoffmann, Ashcroft, and
  Hemley]{Liu:2017-La-Y}
Liu,~H.; Naumov,~I.~I.; Hoffmann,~R.; Ashcroft,~N.~W.; Hemley,~R.~J. Potential
  High-T$_c$ Superconducting Lanthanum and Yttrium Hydrides at High Pressure.
  \emph{Proc. Natl. Acad. Sci. U.S.A.} \textbf{2017}, \emph{114},
  6990--6995\relax
\mciteBstWouldAddEndPuncttrue
\mciteSetBstMidEndSepPunct{\mcitedefaultmidpunct}
{\mcitedefaultendpunct}{\mcitedefaultseppunct}\relax
\EndOfBibitem
\bibitem[P\'epin et~al.(2015)P\'epin, Loubeyre, Occelli, and
  Dumas]{Pepin:2015a}
P\'epin,~C.; Loubeyre,~P.; Occelli,~F.; Dumas,~P. Synthesis of Lithium
  Polyhydrides above 130~GPa at 300~K. \emph{Proc. Natl. Acad. Sci. U.S.A.}
  \textbf{2015}, \emph{112}, 7673--7676\relax
\mciteBstWouldAddEndPuncttrue
\mciteSetBstMidEndSepPunct{\mcitedefaultmidpunct}
{\mcitedefaultendpunct}{\mcitedefaultseppunct}\relax
\EndOfBibitem
\bibitem[Struzhkin et~al.(2016)Struzhkin, Kim, Stavrou, Muramatsu, Mao,
  Pickard, Needs, Prakapenka, and Goncharov]{Struzhkin:2016}
Struzhkin,~V.~V.; Kim,~D.~Y.; Stavrou,~E.; Muramatsu,~T.; Mao,~H.~K.;
  Pickard,~C.~J.; Needs,~R.~J.; Prakapenka,~V.~B.; Goncharov,~A.~F. Synthesis
  of Sodium Polyhydrides at High Pressures. \emph{Nat. Commun.} \textbf{2016},
  \emph{7}, 12267 (1--8)\relax
\mciteBstWouldAddEndPuncttrue
\mciteSetBstMidEndSepPunct{\mcitedefaultmidpunct}
{\mcitedefaultendpunct}{\mcitedefaultseppunct}\relax
\EndOfBibitem
\bibitem[Geballe et~al.(2018)Geballe, Liu, Mishra, Ahart, Somayazulu, Meng,
  Baldini, and Hemley]{Geballe:2018a}
Geballe,~Z.~M.; Liu,~H.; Mishra,~A.~K.; Ahart,~M.; Somayazulu,~M.; Meng,~Y.;
  Baldini,~M.; Hemley,~R.~J. Synthesis and Stability of Lanthanum
  Superhydrides. \emph{Angew. Chem. Int. Ed.} \textbf{2018}, \emph{57},
  688--692\relax
\mciteBstWouldAddEndPuncttrue
\mciteSetBstMidEndSepPunct{\mcitedefaultmidpunct}
{\mcitedefaultendpunct}{\mcitedefaultseppunct}\relax
\EndOfBibitem
\bibitem[Fukai et~al.(1982)Fukai, Fukizawa, Watanabe, and Amano]{Fukai:1982}
Fukai,~Y.; Fukizawa,~A.; Watanabe,~K.; Amano,~M. Hydrogen in Iron--Its Enhanced
  Dissolution under Pressure and Stabilization of the $\gamma$ Phase.
  \emph{Jpn. J. Appl. Phys.} \textbf{1982}, \emph{21}, L318--L320\relax
\mciteBstWouldAddEndPuncttrue
\mciteSetBstMidEndSepPunct{\mcitedefaultmidpunct}
{\mcitedefaultendpunct}{\mcitedefaultseppunct}\relax
\EndOfBibitem
\bibitem[Badding et~al.(1991)Badding, Hemley, and Mao]{Badding:1991}
Badding,~J.~V.; Hemley,~R.~J.; Mao,~H.~K. High-Pressure Chemistry of Hydrogen
  in Metals: \textit{In Situ} Study of Iron Hydride. \emph{Science}
  \textbf{1991}, \emph{253}, 421--424\relax
\mciteBstWouldAddEndPuncttrue
\mciteSetBstMidEndSepPunct{\mcitedefaultmidpunct}
{\mcitedefaultendpunct}{\mcitedefaultseppunct}\relax
\EndOfBibitem
\bibitem[Narygina et~al.(2011)Narygina, Dubrovinsky, McCammon, Kurnosov,
  Kantor, Prakapenka, and Dubrovinskaia]{Narygina:2011}
Narygina,~O.; Dubrovinsky,~L.~S.; McCammon,~C.~A.; Kurnosov,~A.; Kantor,~I.~Y.;
  Prakapenka,~V.~B.; Dubrovinskaia,~N.~A. X-Ray Diffraction and M{\"o}ssbauer
  Spectroscopy Study of fcc Iron Hydride FeH at High Pressures and Implications
  for the Composition of the Earth's Core. \emph{Earth Planet. Sci. Lett.}
  \textbf{2011}, \emph{307}, 409--414\relax
\mciteBstWouldAddEndPuncttrue
\mciteSetBstMidEndSepPunct{\mcitedefaultmidpunct}
{\mcitedefaultendpunct}{\mcitedefaultseppunct}\relax
\EndOfBibitem
\bibitem[Isaev et~al.(2007)Isaev, Skorodumova, Ahuja, Vekilov, and
  Johansson]{Isaev:2007}
Isaev,~E.~I.; Skorodumova,~N.~V.; Ahuja,~R.; Vekilov,~Y.~K.; Johansson,~B.
  Dynamical Stability of Fe-H in the Earth's Mantle and Core Regions.
  \emph{Proc. Natl. Acad. Sci. U.S.A.} \textbf{2007}, \emph{104},
  9168--9171\relax
\mciteBstWouldAddEndPuncttrue
\mciteSetBstMidEndSepPunct{\mcitedefaultmidpunct}
{\mcitedefaultendpunct}{\mcitedefaultseppunct}\relax
\EndOfBibitem
\bibitem[Bazhanova et~al.(2012)Bazhanova, Oganov, and Gianola]{Bazhanova:2012a}
Bazhanova,~Z.~G.; Oganov,~A.~R.; Gianola,~O. Fe-C and Fe-H Systems at Pressures
  of the Earth's Inner Core. \emph{Phys. Usp.} \textbf{2012}, \emph{55},
  489--497\relax
\mciteBstWouldAddEndPuncttrue
\mciteSetBstMidEndSepPunct{\mcitedefaultmidpunct}
{\mcitedefaultendpunct}{\mcitedefaultseppunct}\relax
\EndOfBibitem
\bibitem[Li et~al.(2017)Li, Wang, Du, Zhou, Ma, and Liu]{Li:2017a}
Li,~F.; Wang,~D.; Du,~H.; Zhou,~D.; Ma,~Y.; Liu,~Y. Structural Evolution of
  FeH$_4$ Under High Pressure. \emph{RSC Adv.} \textbf{2017}, \emph{7},
  12570--12575\relax
\mciteBstWouldAddEndPuncttrue
\mciteSetBstMidEndSepPunct{\mcitedefaultmidpunct}
{\mcitedefaultendpunct}{\mcitedefaultseppunct}\relax
\EndOfBibitem
\bibitem[Kvashnin et~al.(2018)Kvashnin, Kruglov, Semenok, and
  Oganov]{Kvashnin:2018a}
Kvashnin,~A.~G.; Kruglov,~I.~A.; Semenok,~D.~V.; Oganov,~A.~R. Iron
  Superhydrides FeH$_5$ and FeH$_6$: Stability, Electronic Properties, and
  Superconductivity. \emph{J. Phys. Chem. C} \textbf{2018}, \emph{122},
  4731--4736\relax
\mciteBstWouldAddEndPuncttrue
\mciteSetBstMidEndSepPunct{\mcitedefaultmidpunct}
{\mcitedefaultendpunct}{\mcitedefaultseppunct}\relax
\EndOfBibitem
\bibitem[P\'epin et~al.(2014)P\'epin, Dewaele, Geneste, and
  Loubeyre]{Pepin:2014}
P\'epin,~C.~M.; Dewaele,~A.; Geneste,~G.; Loubeyre,~P. New Iron Hydrides under
  High Pressure. \emph{Phys. Rev. Lett.} \textbf{2014}, \emph{113}, 265504
  (1--5)\relax
\mciteBstWouldAddEndPuncttrue
\mciteSetBstMidEndSepPunct{\mcitedefaultmidpunct}
{\mcitedefaultendpunct}{\mcitedefaultseppunct}\relax
\EndOfBibitem
\bibitem[Majumdar et~al.(2017)Majumdar, Tse, Wu, and Yao]{Yansun-FeH-2017}
Majumdar,~A.; Tse,~J.~S.; Wu,~M.; Yao,~Y. Superconductivity in FeH$_5$.
  \emph{Phy. Rev. B.} \textbf{2017}, \emph{96}, 201107(R) (1--4)\relax
\mciteBstWouldAddEndPuncttrue
\mciteSetBstMidEndSepPunct{\mcitedefaultmidpunct}
{\mcitedefaultendpunct}{\mcitedefaultseppunct}\relax
\EndOfBibitem
\bibitem[Heil et~al.(2018)Heil, Bachelet, and Boeri]{Heil:2018a}
Heil,~C.; Bachelet,~G.~B.; Boeri,~L. Absence of Superconductivity in Iron
  Polyhydrides at High Pressures. \emph{Phys. Rev. B} \textbf{2018}, \emph{97},
  214510 (1--8)\relax
\mciteBstWouldAddEndPuncttrue
\mciteSetBstMidEndSepPunct{\mcitedefaultmidpunct}
{\mcitedefaultendpunct}{\mcitedefaultseppunct}\relax
\EndOfBibitem
\bibitem[Lonie and Zurek(2011)Lonie, and Zurek]{Zurek:2011a}
Lonie,~D.~C.; Zurek,~E. \textsc{XtalOpt}: An Open-Source Evolutionary Algorithm
  for Crystal Structure Prediction. \emph{Comput. Phys. Commun.} \textbf{2011},
  \emph{182}, 372--387\relax
\mciteBstWouldAddEndPuncttrue
\mciteSetBstMidEndSepPunct{\mcitedefaultmidpunct}
{\mcitedefaultendpunct}{\mcitedefaultseppunct}\relax
\EndOfBibitem
\bibitem[Avery et~al.(2017)Avery, Falls, and Zurek]{Avery:2017-xtalopt}
Avery,~P.; Falls,~Z.; Zurek,~E. \textsc{XtalOpt} Version r10: An Open--Source
  Evolutionary Algorithm for Crystal Structure Prediction. \emph{Comput. Phys.
  Commun.} \textbf{2017}, \emph{217}, 210--211\relax
\mciteBstWouldAddEndPuncttrue
\mciteSetBstMidEndSepPunct{\mcitedefaultmidpunct}
{\mcitedefaultendpunct}{\mcitedefaultseppunct}\relax
\EndOfBibitem
\bibitem[Avery et~al.(2018)Avery, Falls, and Zurek]{Zurek:2017k}
Avery,~P.; Falls,~Z.; Zurek,~E. \textsc{XtalOpt} Version r11: An Open-Source
  Evolutionary Algorithm for Crystal Structure Prediction. \emph{Comput. Phys.
  Commun.} \textbf{2018}, \emph{222}, 418--419\relax
\mciteBstWouldAddEndPuncttrue
\mciteSetBstMidEndSepPunct{\mcitedefaultmidpunct}
{\mcitedefaultendpunct}{\mcitedefaultseppunct}\relax
\EndOfBibitem
\bibitem[Wang et~al.(2012)Wang, Lv, Zhu, and Ma]{wang2012calypso}
Wang,~Y.; Lv,~J.; Zhu,~L.; Ma,~Y. CALYPSO: A Method for Crystal Structure
  Prediction. \emph{Comput. Phys. Commun.} \textbf{2012}, \emph{183},
  2063--2070\relax
\mciteBstWouldAddEndPuncttrue
\mciteSetBstMidEndSepPunct{\mcitedefaultmidpunct}
{\mcitedefaultendpunct}{\mcitedefaultseppunct}\relax
\EndOfBibitem
\bibitem[Lonie and Zurek(2012)Lonie, and Zurek]{Zurek:2011i}
Lonie,~D.~C.; Zurek,~E. Identifying Duplicate Crystal Structures:
  \textsc{XtalComp}, an Open--Source Solution. \emph{Comput. Phys. Commun.}
  \textbf{2012}, \emph{183}, 690--697\relax
\mciteBstWouldAddEndPuncttrue
\mciteSetBstMidEndSepPunct{\mcitedefaultmidpunct}
{\mcitedefaultendpunct}{\mcitedefaultseppunct}\relax
\EndOfBibitem
\bibitem[Kresse and Hafner(1993)Kresse, and Hafner]{Kresse:1993a}
Kresse,~G.; Hafner,~J. \textit{Ab Initio} Molecular Dynamics for Liquid Metals.
  \emph{Phys. Rev. B.} \textbf{1993}, \emph{47}, 558(R)--561(R)\relax
\mciteBstWouldAddEndPuncttrue
\mciteSetBstMidEndSepPunct{\mcitedefaultmidpunct}
{\mcitedefaultendpunct}{\mcitedefaultseppunct}\relax
\EndOfBibitem
\bibitem[Kresse and Joubert(1999)Kresse, and Joubert]{Kresse:1999a}
Kresse,~G.; Joubert,~D. From Ultrasoft Pseudopotentials to the Projector
  Augmented-Wave Method. \emph{Phys. Rev. B.} \textbf{1999}, \emph{59},
  1758--1775\relax
\mciteBstWouldAddEndPuncttrue
\mciteSetBstMidEndSepPunct{\mcitedefaultmidpunct}
{\mcitedefaultendpunct}{\mcitedefaultseppunct}\relax
\EndOfBibitem
\bibitem[Perdew et~al.(1996)Perdew, Burke, and Ernzerhof]{Perdew:1996a}
Perdew,~J.~P.; Burke,~K.; Ernzerhof,~M. Generalized Gradient Approximation Made
  Simple. \emph{Phys. Rev. Lett.} \textbf{1996}, \emph{77}, 3865--3868\relax
\mciteBstWouldAddEndPuncttrue
\mciteSetBstMidEndSepPunct{\mcitedefaultmidpunct}
{\mcitedefaultendpunct}{\mcitedefaultseppunct}\relax
\EndOfBibitem
\bibitem[Bl\"ochl(1994)]{Blochl:1994a}
Bl\"ochl,~P.~E. Projector Augmented-Wave Method. \emph{Phys. Rev. B.}
  \textbf{1994}, \emph{50}, 17953--17979\relax
\mciteBstWouldAddEndPuncttrue
\mciteSetBstMidEndSepPunct{\mcitedefaultmidpunct}
{\mcitedefaultendpunct}{\mcitedefaultseppunct}\relax
\EndOfBibitem
\bibitem[Zhang et~al.(2018)Zhang, Lin, Wang, Yang, Bergara, and
  Ma]{Zhang:2018a}
Zhang,~S.; Lin,~J.; Wang,~Y.; Yang,~G.; Bergara,~A.; Ma,~Y. Nonmetallic FeH$_6$
  under High Pressure. \emph{J. Phys. Chem. C} \textbf{2018}, \emph{122},
  12022--12028\relax
\mciteBstWouldAddEndPuncttrue
\mciteSetBstMidEndSepPunct{\mcitedefaultmidpunct}
{\mcitedefaultendpunct}{\mcitedefaultseppunct}\relax
\EndOfBibitem
\bibitem[Parlinski et~al.(1997)Parlinski, Li, and Kawazoe]{parlinski1997}
Parlinski,~K.; Li,~Z.~Q.; Kawazoe,~Y. First-Principles Determination of the
  Soft Mode in Cubic ZrO$_2$. \emph{Phys. Rev. Lett.} \textbf{1997}, \emph{78},
  4063--4066\relax
\mciteBstWouldAddEndPuncttrue
\mciteSetBstMidEndSepPunct{\mcitedefaultmidpunct}
{\mcitedefaultendpunct}{\mcitedefaultseppunct}\relax
\EndOfBibitem
\bibitem[Chaput et~al.(2011)Chaput, Togo, Tanaka, and Hug]{chaput2011phonon}
Chaput,~L.; Togo,~A.; Tanaka,~I.; Hug,~G. Phonon-Phonon Interactions in
  Transition Metals. \emph{Phys. Rev. B.} \textbf{2011}, \emph{84}, 094302
  (1--6)\relax
\mciteBstWouldAddEndPuncttrue
\mciteSetBstMidEndSepPunct{\mcitedefaultmidpunct}
{\mcitedefaultendpunct}{\mcitedefaultseppunct}\relax
\EndOfBibitem
\bibitem[Gonze and Lee(1997)Gonze, and Lee]{gonze1997dynamical}
Gonze,~X.; Lee,~C. Dynamical Matrices, Born Effective Charges, Dielectric
  Permittivity Tensors, and Interatomic Force Constants From Density-Functional
  Perturbation Theory. \emph{Phys. Rev. B.} \textbf{1997}, \emph{55},
  10355--10368\relax
\mciteBstWouldAddEndPuncttrue
\mciteSetBstMidEndSepPunct{\mcitedefaultmidpunct}
{\mcitedefaultendpunct}{\mcitedefaultseppunct}\relax
\EndOfBibitem
\bibitem[Togo et~al.(2008)Togo, Oba, and Tanaka]{phonopy}
Togo,~A.; Oba,~F.; Tanaka,~I. First-Principles Calculations of the Ferroelastic
  Transition Between Rutile-Type and CaCl$_2$-type SiO$_2$ at High Pressures.
  \emph{Phys. Rev. B.} \textbf{2008}, \emph{78}, 134106 (1--9)\relax
\mciteBstWouldAddEndPuncttrue
\mciteSetBstMidEndSepPunct{\mcitedefaultmidpunct}
{\mcitedefaultendpunct}{\mcitedefaultseppunct}\relax
\EndOfBibitem
\bibitem[Giannozzi et~al.(2009)Giannozzi, Baroni, Bonini, Calandra, Car,
  Cavazzoni, Ceresoli, Chiarotti, Cococcioni, Dabo, Corso, de~Gironcoli,
  Fabris, Fratesi, Gebauer, Gerstmann, Gougoussis, Kokalj, Lazzeri,
  Martin-Samos, Marzari, Mauri, Mazzarello, Paolini, Pasquarello, Paulatto,
  Sbraccia, Scandolo, Sclauzero, Seitsonen, Smogunov, Umari, and
  Wentzcovitch]{giannozzi2009pwscf}
Giannozzi,~P.; Baroni,~S.; Bonini,~N.; Calandra,~M.; Car,~R.; Cavazzoni,~C.;
  Ceresoli,~D.; Chiarotti,~G.~L.; Cococcioni,~M.; Dabo,~I. et~al.  QUANTUM
  ESPRESSO: A Modular and Open-Source Software Project for Quantum Simulations
  of Materials. \emph{J. Phys.: Condens. Matter} \textbf{2009}, \emph{21},
  395502 (1--19)\relax
\mciteBstWouldAddEndPuncttrue
\mciteSetBstMidEndSepPunct{\mcitedefaultmidpunct}
{\mcitedefaultendpunct}{\mcitedefaultseppunct}\relax
\EndOfBibitem
\bibitem[Vanderbilt(1990)]{Vanderbilt:1990}
Vanderbilt,~D. Soft Self-Consistent Pseudopotentials in a Generalized
  Eigenvalue Formalism. \emph{Phys. Rev. B.} \textbf{1990}, \emph{41},
  7892(R)--7895(R)\relax
\mciteBstWouldAddEndPuncttrue
\mciteSetBstMidEndSepPunct{\mcitedefaultmidpunct}
{\mcitedefaultendpunct}{\mcitedefaultseppunct}\relax
\EndOfBibitem
\bibitem[Troullier and Martins(1991)Troullier, and Martins]{Troullier:1991}
Troullier,~N.; Martins,~J.~L. Efficient Pseudopotentials for Plane-Wave
  Calculations. \emph{Phys. Rev. B.} \textbf{1991}, \emph{43}, 1993--2006\relax
\mciteBstWouldAddEndPuncttrue
\mciteSetBstMidEndSepPunct{\mcitedefaultmidpunct}
{\mcitedefaultendpunct}{\mcitedefaultseppunct}\relax
\EndOfBibitem
\bibitem[ato()]{atomic}
\url{https://sites.google.com/site/dceresoli/pseudopotentials}, Accessed July
  2018\relax
\mciteBstWouldAddEndPuncttrue
\mciteSetBstMidEndSepPunct{\mcitedefaultmidpunct}
{\mcitedefaultendpunct}{\mcitedefaultseppunct}\relax
\EndOfBibitem
\bibitem[Allen and Dynes(1975)Allen, and Dynes]{Allen:1975}
Allen,~P.~B.; Dynes,~R.~C. Transition Temperature of Strong-Coupled
  Superconductors Reanalyzed. \emph{Phys. Rev. B.} \textbf{1975}, \emph{12},
  905--922\relax
\mciteBstWouldAddEndPuncttrue
\mciteSetBstMidEndSepPunct{\mcitedefaultmidpunct}
{\mcitedefaultendpunct}{\mcitedefaultseppunct}\relax
\EndOfBibitem
\bibitem[Pickard and Needs(2007)Pickard, and Needs]{Pickard-H-2007}
Pickard,~C.~J.; Needs,~R.~J. Structure of Phase III of Solid Hydrogen.
  \emph{Nat. Phys.} \textbf{2007}, \emph{3}, 473--476\relax
\mciteBstWouldAddEndPuncttrue
\mciteSetBstMidEndSepPunct{\mcitedefaultmidpunct}
{\mcitedefaultendpunct}{\mcitedefaultseppunct}\relax
\EndOfBibitem
\bibitem[S{\"o}derlind et~al.(1996)S{\"o}derlind, Moriarty, and Wills]{Fe-Ref}
S{\"o}derlind,~P.; Moriarty,~J.~A.; Wills,~J.~M. First-Principles Theory of
  Iron up to Earth--Core Pressures: Structural, Vibrational, and Elastic
  Properties. \emph{Phys. Rev. B} \textbf{1996}, \emph{53}, 14063--14072\relax
\mciteBstWouldAddEndPuncttrue
\mciteSetBstMidEndSepPunct{\mcitedefaultmidpunct}
{\mcitedefaultendpunct}{\mcitedefaultseppunct}\relax
\EndOfBibitem
\bibitem[Mishra et~al.(2018)Mishra, Muramatsu, Liu, Geballe, Somayazulu, Ahart, Baldini, Meng, Zurek, and Hemley]{Zurek:2018c}
Mishra,~A.~K.; Muramatsu,~T.; Liu,~H.; Geballe,~Z.~M.; Somayazulu,~M.; Ahart,~M.; Baldini,~M.; Meng,~Y.; Zurek,~E.; Hemley,~R.~J. New Calcium Hydrides with Mixed Atomic and Molecular Hydrogen. \emph{J. Phys. Chem. C} \textbf{2018}, \emph{122},
  19370--19378\relax
\mciteBstWouldAddEndPuncttrue
\mciteSetBstMidEndSepPunct{\mcitedefaultmidpunct}
{\mcitedefaultendpunct}{\mcitedefaultseppunct}\relax
\EndOfBibitem
\bibitem[Drozdov et~al.(2015)Drozdov, Eremets, and Troyan]{Drozdov:2015-P}
Drozdov,~A.~P.; Eremets,~M.~I.; Troyan,~I.~A. Superconductivity above 100~K in
  PH$_3$ at High Pressures. \emph{arXiv preprint} \textbf{2015},
  arXiv:1508.06224\relax
\mciteBstWouldAddEndPuncttrue
\mciteSetBstMidEndSepPunct{\mcitedefaultmidpunct}
{\mcitedefaultendpunct}{\mcitedefaultseppunct}\relax
\EndOfBibitem
\bibitem[Shamp et~al.(2016)Shamp, Terpstra, Bi, Falls, Avery, and
  Zurek]{Zurek:2015j}
Shamp,~A.; Terpstra,~T.; Bi,~T.; Falls,~Z.; Avery,~P.; Zurek,~E. Decomposition
  Products of Phosphine Under Pressure: PH2 Stable and Superconducting?
  \emph{J. Am. Chem. Soc.} \textbf{2016}, \emph{138}, 1884--1892\relax
\mciteBstWouldAddEndPuncttrue
\mciteSetBstMidEndSepPunct{\mcitedefaultmidpunct}
{\mcitedefaultendpunct}{\mcitedefaultseppunct}\relax
\EndOfBibitem
\bibitem[Bi et~al.(2017)Bi, Miller, Shamp, Zurek, E.]{Zurek:2017c}
Bi,~T. and Miller,~D.~P.; Shamp,~A.; Zurek,~E. Superconducting Phases of Phosphorus Hydride Under Pressure: Stabilization via Mobile Molecular Hydrogen.
  \emph{Angew. Chem. Int. Ed.} \textbf{2017}, \emph{56}, 10192--10195\relax
\mciteBstWouldAddEndPuncttrue
\mciteSetBstMidEndSepPunct{\mcitedefaultmidpunct}
{\mcitedefaultendpunct}{\mcitedefaultseppunct}\relax
\EndOfBibitem
\bibitem[Cudazzo et~al.(2008)Cudazzo, Profeta, Sanna, Floris, Continenza,
  Massidda, and Gross]{Cudazzo:2008a}
Cudazzo,~P.; Profeta,~G.; Sanna,~A.; Floris,~A.; Continenza,~A.; Massidda,~S.;
  Gross,~E. K.~U. \textit{Ab Initio} Description of High--Temperature
  Superconductivity in Dense Molecular Hydrogen. \emph{Phys. Rev. Lett.}
  \textbf{2008}, \emph{100}, 257001 (1--4)\relax
\mciteBstWouldAddEndPuncttrue
\mciteSetBstMidEndSepPunct{\mcitedefaultmidpunct}
{\mcitedefaultendpunct}{\mcitedefaultseppunct}\relax
\EndOfBibitem
\bibitem[Ponc\'e et~al.(2016)Ponc\'e, Margine, Verdi, and Giustino]{EPW}
Ponc\'e,~S.; Margine,~E.~R.; Verdi,~C.; Giustino,~F. EPW: Electron-Phonon
  Coupling, Transport and Superconducting Properties Using Maximally Localized
  Wannier Functions. \emph{Comput. Phys. Commun.} \textbf{2016}, \emph{209},
  116--133\relax
\mciteBstWouldAddEndPuncttrue
\mciteSetBstMidEndSepPunct{\mcitedefaultmidpunct}
{\mcitedefaultendpunct}{\mcitedefaultseppunct}\relax
\EndOfBibitem
\bibitem[Wierzbowska et~al.(2006)Wierzbowska, de~Gironcoli, and
  Giannozzi]{Wierzbowska}
Wierzbowska,~M.; de~Gironcoli,~S.; Giannozzi,~P. Origins of Low- and High-
  Pressure Discontinuities of T$_c$ in Niobium. \emph{arXiv preprint} \textbf{2006},
  arXiv:cond-mat/0504077v2\relax
\mciteBstWouldAddEndPuncttrue
\mciteSetBstMidEndSepPunct{\mcitedefaultmidpunct}
{\mcitedefaultendpunct}{\mcitedefaultseppunct}\relax
\EndOfBibitem
\bibitem[McMahon and Ceperley(2011)McMahon, and Ceperley]{Mcmahon-2011}
McMahon,~J.~M.; Ceperley,~D.~M. High-Temperature Superconductivity in Atomic
  Metallic Hydrogen. \emph{Phy. Rev. B.} \textbf{2011}, \emph{84}, 144515
  (1--8)\relax
\mciteBstWouldAddEndPuncttrue
\mciteSetBstMidEndSepPunct{\mcitedefaultmidpunct}
{\mcitedefaultendpunct}{\mcitedefaultseppunct}\relax
\EndOfBibitem
\bibitem[Wang et~al.(2018)Wang, Duan, Yu, Xie, Huang, Ma, Tian, Li, Liu, and
  Cui]{CoH-Cui-2018}
Wang,~L.; Duan,~D.; Yu,~H.; Xie,~H.; Huang,~X.; Ma,~Y.; Tian,~F.; Li,~D.;
  Liu,~B.; Cui,~T. High-Pressure Formation of Cobalt Polyhydrides: A
  First-Principle Study. \emph{Inorg. Chem.} \textbf{2018}, \emph{57},
  181--186\relax
\mciteBstWouldAddEndPuncttrue
\mciteSetBstMidEndSepPunct{\mcitedefaultmidpunct}
{\mcitedefaultendpunct}{\mcitedefaultseppunct}\relax
\EndOfBibitem
\bibitem[Liu et~al.(2016)Liu, Duan, Tian, Wang, Ma, Li, Huang, Liu, and
  Cui]{Liu:2015c}
Liu,~Y.; Duan,~D.; Tian,~F.; Wang,~C.; Ma,~Y.; Li,~D.; Huang,~X.; Liu,~B.;
  Cui,~T. Stability of Properties of the Ru-H System at High Pressure.
  \emph{Phys. Chem. Chem. Phys.} \textbf{2016}, \emph{18}, 1516--1520\relax
\mciteBstWouldAddEndPuncttrue
\mciteSetBstMidEndSepPunct{\mcitedefaultmidpunct}
{\mcitedefaultendpunct}{\mcitedefaultseppunct}\relax
\EndOfBibitem
\bibitem[Liu et~al.(2015)Liu, Duan, Huang, Tian, Li, Sha, Wang, Zhang, Yang,
  Liu, and Cui]{Liu:2015b}
Liu,~Y.; Duan,~D.; Huang,~X.; Tian,~F.; Li,~D.; Sha,~X.; Wang,~C.; Zhang,~H.;
  Yang,~T.; Liu,~B. et~al.  Structures and Properties of Osmium Hydrides under
  Pressure from First Principle Calculation. \emph{J. Phys. Chem. C.}
  \textbf{2015}, \emph{119}, 15905--15911\relax
\mciteBstWouldAddEndPuncttrue
\mciteSetBstMidEndSepPunct{\mcitedefaultmidpunct}
{\mcitedefaultendpunct}{\mcitedefaultseppunct}\relax
\EndOfBibitem
\bibitem[Kohlhaas et~al.(1967)Kohlhaas, Dunner, and
  Schmitz-Prange]{Kohlhaas:1967}
Kohlhaas,~R.; Dunner,~P.; Schmitz-Prange,~N. The Temperature Dependance of the
  Lattice Parameters of Iron, Cobalt, and Nickel in the High Temperature Range.
  \emph{Z. Angew. Phys.} \textbf{1967}, \emph{23}, 245--249\relax
\mciteBstWouldAddEndPuncttrue
\mciteSetBstMidEndSepPunct{\mcitedefaultmidpunct}
{\mcitedefaultendpunct}{\mcitedefaultseppunct}\relax
\EndOfBibitem
\bibitem[Hooper and Zurek(2012)Hooper, and Zurek]{Zurek:2011h}
Hooper,~J.; Zurek,~E. Rubidium Polyhydrides Under Pressure: Emergence of the
  Linear H$_3^-$ Species. \emph{Chem. Eur. J.} \textbf{2012}, \emph{18},
  5013--5021\relax
\mciteBstWouldAddEndPuncttrue
\mciteSetBstMidEndSepPunct{\mcitedefaultmidpunct}
{\mcitedefaultendpunct}{\mcitedefaultseppunct}\relax
\EndOfBibitem
\bibitem[Zhong et~al.(2016)Zhong, Wang, Zhang, Liu, Zhang, Song, Yang, Zhang,
  and Ma]{Zhong:2016a}
Zhong,~X.; Wang,~H.; Zhang,~J.; Liu,~H.; Zhang,~S.; Song,~H.~F.; Yang,~G.;
  Zhang,~L.; Ma,~Y. Tellurium Hydrides at High Pressures: High-Temperature
  Superconductors. \emph{Phys. Rev. Lett.} \textbf{2016}, \emph{116}, 057002
  (1--6)\relax
\mciteBstWouldAddEndPuncttrue
\mciteSetBstMidEndSepPunct{\mcitedefaultmidpunct}
{\mcitedefaultendpunct}{\mcitedefaultseppunct}\relax
\EndOfBibitem
\end{mcitethebibliography}

\providecommand*\mcitethebibliography{\thebibliography}
\csname @ifundefined\endcsname{endmcitethebibliography}
  {\let\endmcitethebibliography\endthebibliography}{}

\newpage

\textbf{TOC Graphic}

\begin{figure*}
\begin{center}
\includegraphics[width=8.25cm]{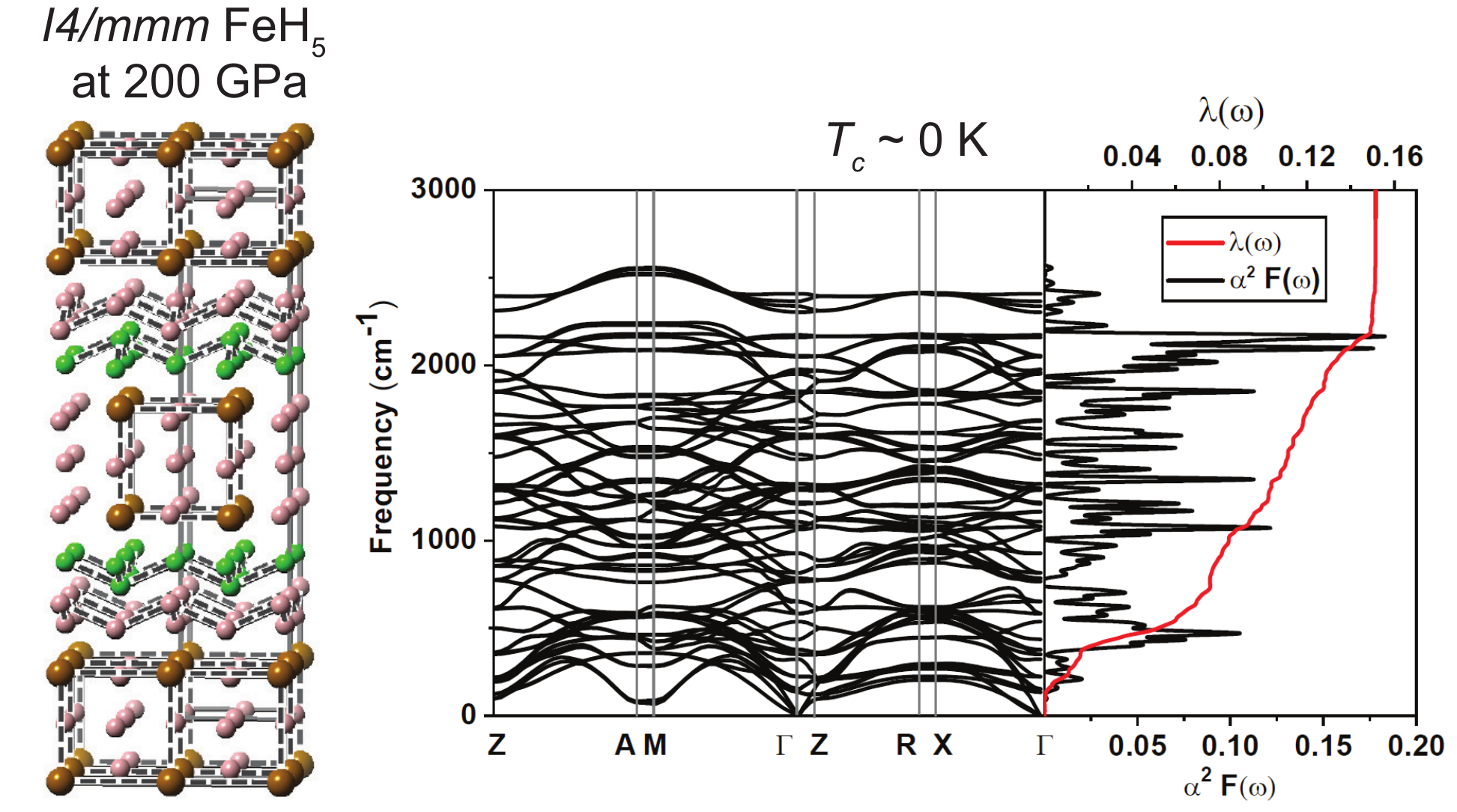}
\end{center} 
\end{figure*}

\end{document}